\documentclass{article}

\usepackage{PRIMEarxiv}
\usepackage{amsmath}
\usepackage[utf8]{inputenc} 
\usepackage[T1]{fontenc}    
\usepackage{hyperref}       
\usepackage{url}            
\usepackage{booktabs}       
\usepackage{amsfonts}       
\usepackage{nicefrac}       
\usepackage{microtype}      
\usepackage{lipsum}
\usepackage{fancyhdr}       
\usepackage{graphicx}       
\graphicspath{{media/}}

\title{The AI Collaborator: Bridging Human-AI Interaction in Educational and Professional Settings}

\author{
  Mohammad Amin Samadi \\
  University of California, Irvine \\
  Irvine, CA, USA \\
  \texttt{masamadi@uci.edu}
  \and
  Spencer JaQuay \\
  University of California, Irvine \\
  Irvine, CA, USA \\
  \texttt{sjaquay@uci.edu}
  \and
  Jing Gu \\
  University of California, Irvine \\
  Irvine, CA, USA \\
  \texttt{jgu12@uci.edu}
  \and
  Nia Nixon \\
  University of California, Irvine \\
  Irvine, CA, USA \\
  \texttt{dowelln@uci.edu}
}

\begin{document}
\maketitle

\begin{abstract}
AI Collaborator, powered by OpenAI's GPT-4, is a groundbreaking tool designed for human-AI collaboration research. Its standout feature is the ability for researchers to create customized AI personas for diverse experimental setups using a user-friendly interface. This functionality is essential for simulating various interpersonal dynamics in team settings. AI Collaborator excels in mimicking different team behaviors, enabled by its advanced memory system and a sophisticated personality framework. Researchers can tailor AI personas along a spectrum from dominant to cooperative, enhancing the study of their impact on team processes. The tool's modular design facilitates integration with digital platforms like Slack, making it versatile for various research scenarios. AI Collaborator is thus a crucial resource for exploring human-AI team dynamics more profoundly.
\end{abstract}


\section{Introduction}

In the rapidly evolving landscape of artificial intelligence, significant advancements are being made, impacting a broad spectrum of fields ranging from Education \cite{becker2018nmc} to road transit \cite{banks2019analysis}. Looking ahead, these advancements are poised to significantly influence the dynamics of team environments. While research on teams only a few years ago highlighted the potential usefulness of AI integration in both research and practical settings, it also acknowledged the limitations of AI technologies in fully mimicking and comprehending the complex aspects of human-team interactions at the time \cite{seeber2020machines}. However, with recent developments in generative AI and Large Language Models i.e., (OpenAI's GPT-4 \cite{openai2023gpt4}, Google's Bard \cite{manyika2023overview} and Gemini \cite{team2023gemini}), we are approaching a level where AI-human teams can collaborate more effectively e.g., \cite{lakhnati2023exploring}. This progression prompts a critical question: How can we harness the evolving capabilities of AI to effectively enhance and integrate it into human-AI team dynamics, particularly in settings where traditional automation tools face limitations?

This shift in AI's role within teams is underpinned by rapid advancements in technology, enhancing the agents' capacity for complex functions within team settings. Human-Autonomy Teaming (HATs) represents a sophisticated aspect of human-machine interaction, focusing on collaborative dynamics between humans and autonomous systems, such as AI-driven agents \cite{adler1997work}. This evolution from traditional tools centered around human-machine interaction to autonomous agents capable of participating in intricate team settings illustrates a deeper integration of AI technologies in collaborative environments \cite{mcneese2018teaming}. However, the full potential of these AI collaboration tools in broader applications remains largely untapped. This is primarily due to persisting limitations in AI's adaptability, decision-making autonomy, and communication abilities \cite{demir2016team,seeber2020machines}. These limitations underscore the need for more sophisticated AI systems that can accurately replicate and adapt to the dynamic nuances of human teamwork, thereby unlocking the true potential of AI-enhanced collaborative environments.

In response to these challenges, the work of Seeber et al. (2019) provides a foundational call to action, raising critical questions about the conditions under which humans accept machines as teammates \cite{seeber2020machines}. Their investigation into factors such as the machine's demeanor and its role in coordination versus creative tasks highlights how these influence the acceptance of AI in team dynamics. They probe into the extent to which human collaborators accept machine input, the impact of different styles of verbal and nonverbal communication on this acceptance, and the nature of trust in machine-generated recommendations.

We also consider the role of 'suspension of disbelief' in AI team interactions. This concept, originating from Samuel Taylor Coleridge in 1817, refers to the willingness of a person to accept as true the premises of a work of fiction, even if they are fantastic or impossible. In the context of AI, this means that if an AI bot exhibits a realistic and coherent personality, team members might interact with it as they would with a human, overlooking its artificial nature \cite{prada2010introducing}. This psychological phenomenon is critical for the AI's integration into team dynamics, influencing how team members perceive and respond to its actions and suggestions. Recent findings by Zhang et al. (2023) lend support to this concept, revealing that AI teammates' proactive communication with humans could facilitate the development of human trust and situational awareness. Conversely, AI that lacks such proactive communication is often not perceived as a teammate \cite{zhang2023investigating}. This highlights the need for AI's active engagement in communication and its role in fostering a cohesive team environment.

In recent advancements, large language models have demonstrated their capability to encode a wide array of human behaviors \cite{59_brown2020language,60_bommasani2022opportunities}, introducing innovative means for crafting authentic AI personalities. The study "Generative Agents: Interactive Simulacra of Human Behavior" \cite{park2023generative} exemplifies this progress, employing large language models with an integrated memory component. This approach simulates human cognition, enabling agents to exhibit believable social behaviors. Such developments are pivotal to our endeavor in creating an AI bot that not only mirrors human personality but also adapts based on learned experiences and team interactions, thereby fostering more authentic human-like interactions.

These insights have profoundly influenced our approach, compelling us to develop an innovative AI bot designed for collaboration. In this manuscript, we introduce technology tailored to advance the field of HATs. Our focus lies in meticulously replicating two critical aspects of human behavior: the memory system and personality. Our objective is to refine our bot’s capabilities, enabling it to emulate human-like behavior and interactions more accurately. This advancement promises to provide researchers with a bot simulator that not only replicates a human-like collaborative atmosphere but also does so with validity and reliability. Such a technological leap will empower researchers to effectively address complex challenges in team research, such as evaluating the influence of a dominating personality on the sense of inclusion among traditionally marginalized communities and exploring AI teaming dynamics in critical scenarios like first responder decision-making. Importantly, it enables these studies to be conducted at a scale where meaningful conclusions can be robustly drawn. Our work marks a significant step forward in the realm of AI, striving to bridge the gap between artificial and human intelligence in team dynamics.

\section{Our AI Collaborator}

This section outlines the main features of the AI Collaborator, detailing how each component contributes to its overall functionality and effectiveness. Additionally, we will discuss future development plans, emphasizing our commitment to continuous improvement and adaptation in response to emerging challenges, technological trends, and user feedback.

\begin{figure}
    \centering
    \includegraphics[width=0.85\textwidth]{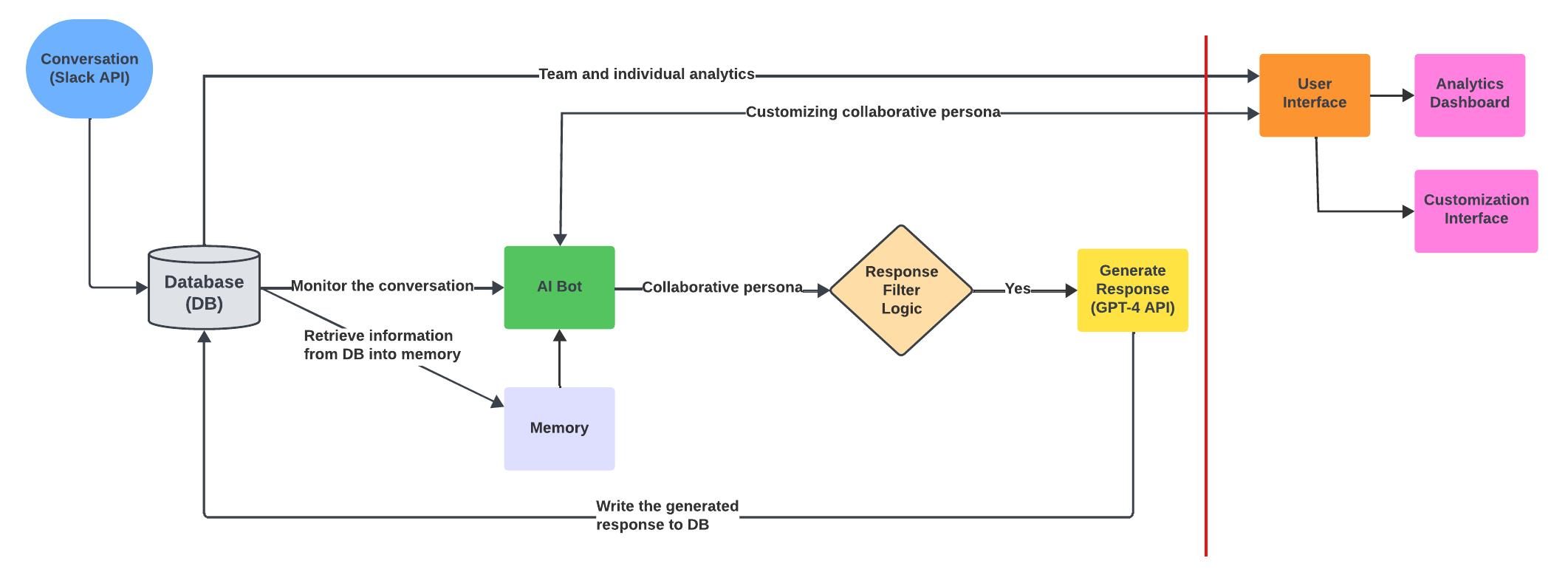}
    \caption{Overview of the general workflow of AI Collaborator, visualizing the main components and how they interact.}
    \label{fig:workflow}
\end{figure}

\subsection{System Architecture}

The AI Collaborator platform is an integration of multiple technologies, combining the Slack collaborative chat environment and its API to capture the ongoing participants-AI dialogue. This conversational data is subsequently processed and stored within a dedicated database. The AI bot interacts with the database, which retrieves the necessary information to formulate responses. The information is then processed through the logic filter which allows for customized response control. These responses are generated using OpenAI's GPT-4 API, specifically employing the \textit{chat.completions} function. Here, a customized persona is incorporated via system role instructions\footnote{https://platform.openai.com/docs/api-reference/chat/create}. On the front-end, the platform features a user interface (UI) that displays team and individual dynamics via a comprehensive dashboard. This UI also provides an interactive space for researchers to initiate tasks, tailor the bot’s behavior and persona, and define team composition requirements. The development trajectory of the UI is elaborated upon in section \ref{sec:user_interface}. Additionally, Figure \ref{fig:workflow} illustrates the overarching workflow of the AI Collaborator platform. 

\subsection{Memory}
In the AI Collaborator, the memory system originally developed by \cite{park2023generative} has been innovatively adapted to enhance conversational AI agents. This adaptation shifts the focus from recording events to leveraging detailed logs of past conversations. These logs include content, context, and nuances of interactions, enabling the AI to construct and retrieve a rich, conversational history.

Key features of this adapted memory system include:

\begin{itemize}
    \item \textbf{Dynamic Response Generation:} The AI plans responses not only based on current inputs but also using insights from prior interactions, ensuring coherent and contextually appropriate conversations.
    \item \textbf{Reflective Memory Synthesis:} This feature enables the AI to summarize and abstract its interactions into reflective memories periodically. By analyzing and integrating insights from past interactions, it forms higher-level concepts and patterns.
    \item \textbf{Context-Aware Retrieval Model:} This model is fine-tuned to prioritize memories based on their relevance to the ongoing conversation, considering factors like recency, relevance, and importance.
\end{itemize}

\subsubsection*{Information Retrieval from the Database:}
To retrieve information from the database to the memory, the system employs a linear combination of the three measures: Recency, Relevance, and Importance. This combination is used to calculate a composite score for each piece of information in the database. The formula for the composite score is as follows:
\begin{equation}
    \text{Composite Score} = \alpha \cdot \text{Recency} + \beta \cdot \text{Relevance} + \gamma \cdot \text{Importance}
\end{equation}
Here, \( \alpha \), \( \beta \), and \( \gamma \) are weights assigned to the Recency, Relevance, and Importance measures, respectively. These weights are determined through optimization to best suit the system's needs. The system then retrieves the top \( k \) pieces of information based on these composite scores. The value of \( k \) is a parameter that can be optimized through testing, to balance between providing sufficient information and maintaining computational efficiency. This method ensures that the most contextually relevant and recent information is presented in response to the current query or conversation.

\subsubsection*{1. Recency:}
The recency of past interactions is managed using a decay function, which diminishes the weight of older conversations in the memory stream. This ensures that more recent interactions are prioritized. The decay function is modeled as an exponential decay, represented by:
\begin{equation}
    W(t) = e^{-\lambda t}
\end{equation}
Here, \( W(t) \) represents the weight of a conversation at time \( t \), and \( \lambda \) is the decay constant that determines the rate at which older conversations lose significance.

\subsubsection*{2. Relevance:}
To measure relevance, the AI employs a semantic similarity model that compares the content of a previous conversation to the current prompt. This can be quantified using the cosine similarity between the vector representations of the previous conversation and the current prompt:
\begin{equation}
    \text{Similarity} = \cos(\theta) = \frac{\mathbf{A} \cdot \mathbf{B}}{\|\mathbf{A}\|\|\mathbf{B}\|}
\end{equation}
In this formula, \( \mathbf{A} \) and \( \mathbf{B} \) are vector representations of the previous conversation and the current prompt, respectively.

\subsubsection*{3. Importance:}
The importance of a message within the overall conversation is determined by its semantic similarity to the rest of the conversation. The importance score for a message can be modeled as:
\begin{equation}
    \text{Importance}(M) = \frac{1}{N-1} \sum_{i=1}^{N} \text{Similarity}(M, M_i)
\end{equation}
Where \( \text{Importance}(M) \) is the importance score of message \( M \), \( N \) is the total number of messages in the conversation, and \( \text{Similarity}(M, M_i) \) is the semantic similarity between message \( M \) and another message \( M_i \) in the conversation.

\subsection{Persona Customization}

\subsubsection{AI Persona Customization}
The AI Collaborator introduces a pioneering feature for AI persona customization. This innovation centers on translating detailed personality traits into specific, actionable AI behaviors and interaction styles. The development process involves:

\begin{enumerate}
    \item \textit{Translation of Personality Traits:} An in-depth analysis of the Big Five personality traits and their sub-components is conducted to identify specific, actionable characteristics. These are then translated into clear, concise descriptors.
    \item \textit{Creation of Prompt Templates:} Descriptors are formulated into flexible yet precise prompt templates designed to elicit corresponding traits in the AI’s responses, adaptable to varying interaction contexts.
\end{enumerate}

\subsubsection{Integration with AI's Response Mechanisms}
The integration of these prompts into the AI's response system is critical for a natural expression of personality traits:

\begin{enumerate}
    \item \textit{Programming and Machine Learning Models:} Specific responses and behavioral filters are developed and employed to ensure the AI's responses seamlessly exhibit the intended personality traits.
    \item \textit{Feedback Loop for Refinement:} A feedback loop is established where the AI learns from user and internal feedback, crucial for refining its understanding and expression of personality traits.
\end{enumerate}

\section{Testing}

The evaluation of the AI Collaborator is designed to be comprehensive, encompassing both technical performance and user experience. Our methodology integrates a dual approach: technical assessment of the AI model, based on its internal specifications and functionalities, and empirical evaluation of user interactions, gauged through structured experiments.

\subsection{Technical Evaluation}

\textbf{Model Specification Variables:}
\begin{enumerate}
    \item \textit{Memory Variables:} We will examine how variations in the AI's memory components influence its interaction quality and performance efficiency.
    \item \textit{Customization Features:} The effectiveness and user-friendliness of the AI's customization capabilities will be assessed in diverse application scenarios.
\end{enumerate}

\textbf{API Feature Variables:}
\begin{enumerate}
    \setcounter{enumi}{2}
    \item \textit{API Configurations:} The study will investigate how different settings within the OpenAI API, such as temperature control and message length limits, affect the quality and appropriateness of the AI's responses.
\end{enumerate}

\subsection{User Experience Evaluation}

User experience will be quantitatively and qualitatively assessed based on the following criteria:

\begin{enumerate}
    \item \textit{Perception of the AI:} User feedback will be collected to evaluate perceptions of the AI's intelligence, responsiveness, and personality traits.
    \item \textit{Task Relevance and Clarity:} The relevance and clarity of tasks assigned or facilitated by the AI will be assessed through user surveys.
    \item \textit{Overall Experience:} The user's overall satisfaction and engagement with the AI will be evaluated.
    \item \textit{Human-Likeness:} A metric will be developed to rate the degree to which interactions with the AI resemble those with a human.
    \item \textit{Conversation Fluidity:} The naturalness and seamlessness of conversations with the AI will be quantitatively measured.
    \item \textit{Task Effectiveness:} The effectiveness of the AI in assisting with task completion or providing helpful support will be evaluated.
\end{enumerate}

\begin{figure}
    \centering
    \includegraphics[width = 0.8\textwidth]{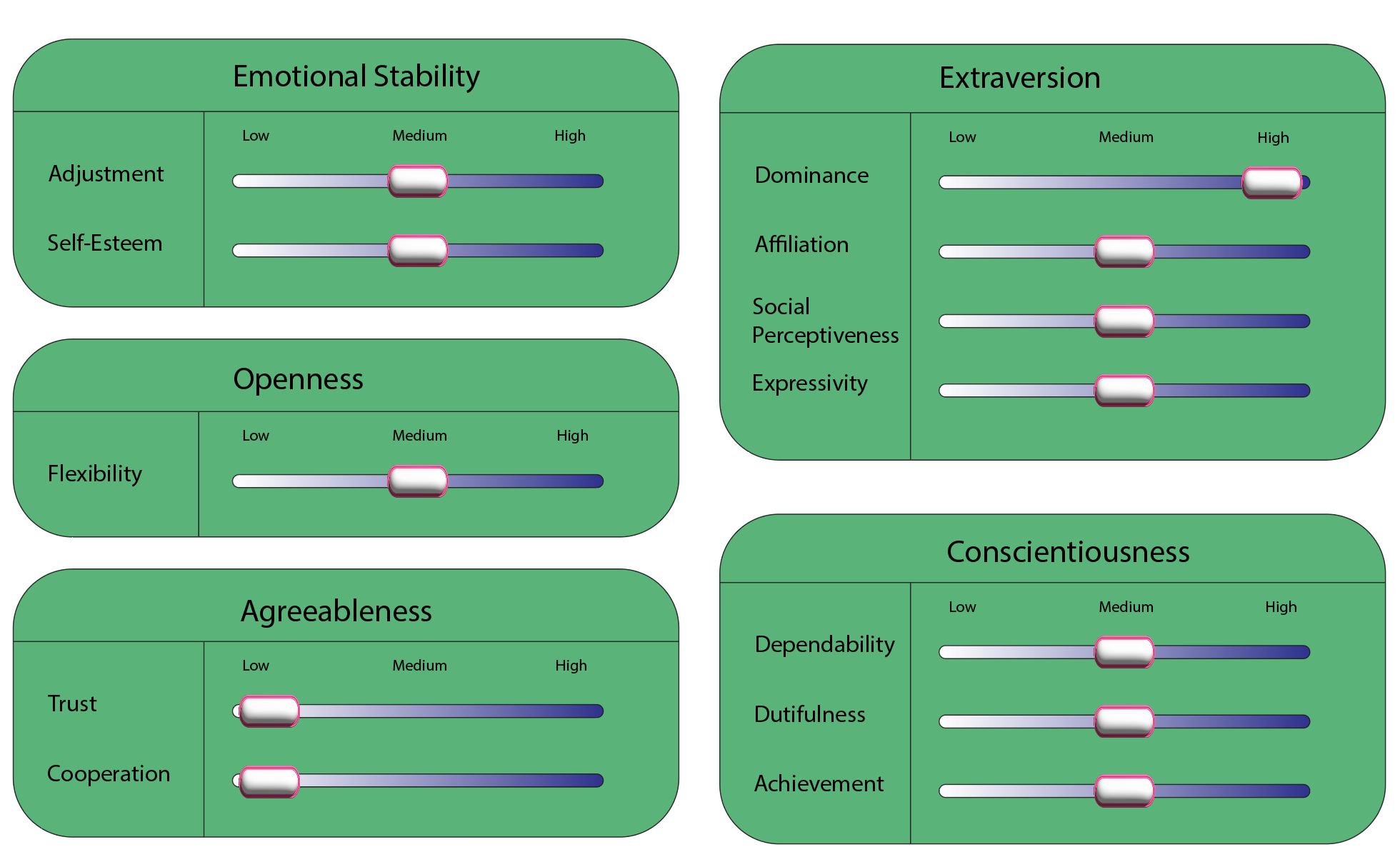}
    \caption{This prototype of the UI illustrates the adjustable expressions of the Big Five personality traits' subcomponents, with three levels of intensity: low, medium, and high. As depicted in this example, the interface presents a collaborative bot characterized by a highly dominant persona. This figure was adapted from the hierarchical model of facets related to teamwork, as detailed in \cite{driskell2006makes}.}
    \label{fig:persona}
\end{figure}

\section{User Interface Developments}
\label{sec:user_interface}

To advance the AI Collaborator, significant enhancements are planned for the UI. The overarching design focus of the UI is to grant researchers complete control over their experimental setups, while simultaneously providing them with detailed analytics. This will enable a comprehensive evaluation of team, individual, and human-AI interaction patterns, thus contributing significantly to collaborative AI research. 

\subsection{Customization Portal for Researchers}
The cornerstone of the UI development is the creation of a Customization Portal specifically designed for researchers. This portal will enable:

\begin{enumerate}
    \item \textit{Scenario-Specific Customization:} Researchers will have the ability to tailor the AI's functionalities and responses to fit specific research scenarios. The initial prototype of the UI settings for customizing the AI persona is illustrated in Figure \ref{fig:persona}.
    \item \textit{Task Specification:} The portal will allow researchers to clearly define the tasks for participating teams, ensuring that each experiment aligns with their research objectives. This feature also will be accompanied by a dropbox to upload relevant contextual information, e.g., external documents.
    \item \textit{Team Composition Control:} Researchers will be able to manipulate team compositions (i.e., team size, gender composition, age) through a matching algorithm to fit research requirements.
    \item \textit{Time Management:} The portal will include functionality to set the duration of tasks, allowing for precise control over the experimental timeline.
\end{enumerate}

\subsection{Dashboard Analytics}
Another key development is the integration of a comprehensive Dashboard Analytics feature. This feature will provide:

\begin{enumerate}
    \item \textit{Team Dynamics:} The dashboard offers insights into interaction patterns between teams and the AI, as well as internal team dynamics.
    \item \textit{Individual Insights:} Tools to analyze individual behaviors and contributions within teams.
\end{enumerate}

\section{Data Collection}

The following data is collected from each collaboration session:

\begin{enumerate}
    \item \textbf{Conversational Data and Metadata:} Team conversations, including content and metadata such as timestamps and speaker identification.

    \item \textbf{Participants' Demographic and Individual Difference Data:} Demographic details are collected via user profiles or questionnaires, including age, gender, and educational background, and individual difference measures, such as sense of belonging, self-efficacy, and personality enabling a nuanced analysis of team composition and diversity.

    \item \textbf{Team and Individual Analytics:} The system processes conversational data and metadata to assess team dynamics and individual contributions.

    \item \textbf{AI's Reflections and Automated Behavioral Tagging:} The AI Collaborator offers periodic reflections on the conversations. Behavioral tagging identifies and codes predefined behaviors within the conversation, enhancing the understanding of team interactions and individual behaviors.

\end{enumerate}

\section{Challenges}

As the development of the AI Collaborator progresses, new challenges continuously emerge and are identified. These evolving challenges are crucial to address for the successful implementation and widespread adoption of the system. The following are some of the key challenges currently recognized:

\begin{enumerate}

    \item \textbf{Handling Conversations with Multiple People}: The AI must be adept at managing interactions involving multiple participants. This involves understanding the context of multi-person conversations, accurately attributing statements to the right individuals, and maintaining coherent and relevant responses within the group dynamic.

    \item \textbf{Proactivity and Controlled Information Sharing}: The AI is required to be proactive in conversations, contributing constructively while ensuring it does not extrapolate or introduce information beyond the scope of the conversation. It should allow the conversation to flow naturally, drawing from its knowledge base only when relevant and appropriate, thus respecting the organic nature of human dialogue.

    \item \textbf{Scalability and Cross-Platform Compatibility}: A primary challenge lies in scaling the AI tool to accommodate a growing number of participants while maintaining performance stability. This must be achieved alongside ensuring that the tool functions seamlessly across various platforms and systems, requiring a flexible and robust architecture.

    \item \textbf{Complexity in AI Personalization}: Customizing AI personas to exhibit a range of human-like behaviors and interactions poses a complex challenge. The AI needs to maintain these personalized characteristics consistently across various scenarios, ensuring the personas remain authentic and relatable to users.

    \item \textbf{Cultural and Ethical Considerations}: Given the diversity of user groups, especially in global settings, the AI must be designed with cultural sensitivities and ethical considerations in mind. Its behavior and responses need to be appropriate, respectful, and inclusive of cultural differences to ensure universal acceptability and effectiveness.
\end{enumerate}

\section{Contributions and Implications}
This study presents a robust platform that offers a novel approach to investigating fundamental questions in human-AI teaming. To achieve this, we developed an innovative memory system and a sophisticated personality framework, which marks a major advancement in creating immersive human-AI experiences. These developments serve as a vital reference for future research in the field, addressing current technological challenges and setting new directions for exploration. Our work establishes a foundational platform for understanding and improving team dynamics, especially in AI-integrated settings.
\bibliographystyle{ACM-Reference-Format}
\bibliography{ref}


\begin{thebibliography}{16}


\ifx \showCODEN    \undefined \def \showCODEN     #1{\unskip}     \fi
\ifx \showDOI      \undefined \def \showDOI       #1{#1}\fi
\ifx \showISBNx    \undefined \def \showISBNx     #1{\unskip}     \fi
\ifx \showISBNxiii \undefined \def \showISBNxiii  #1{\unskip}     \fi
\ifx \showISSN     \undefined \def \showISSN      #1{\unskip}     \fi
\ifx \showLCCN     \undefined \def \showLCCN      #1{\unskip}     \fi
\ifx \shownote     \undefined \def \shownote      #1{#1}          \fi
\ifx \showarticletitle \undefined \def \showarticletitle #1{#1}   \fi
\ifx \showURL      \undefined \def \showURL       {\relax}        \fi
\providecommand\bibfield[2]{#2}
\providecommand\bibinfo[2]{#2}
\providecommand\natexlab[1]{#1}
\providecommand\showeprint[2][]{arXiv:#2}

\bibitem[Adler(1997)]%
        {adler1997work}
\bibfield{author}{\bibinfo{person}{Paul~S Adler}.} \bibinfo{year}{1997}\natexlab{}.
\newblock \showarticletitle{Work organization: From taylorism to teamwork}.
\newblock \bibinfo{journal}{\emph{Perspectives on Work}} \bibinfo{volume}{1}, \bibinfo{number}{1} (\bibinfo{year}{1997}), \bibinfo{pages}{61--65}.
\newblock


\bibitem[Banks and Stanton(2019)]%
        {banks2019analysis}
\bibfield{author}{\bibinfo{person}{Victoria~A Banks} {and} \bibinfo{person}{Neville~A Stanton}.} \bibinfo{year}{2019}\natexlab{}.
\newblock \showarticletitle{Analysis of driver roles: Modelling the changing role of the driver in automated driving systems using EAST}.
\newblock \bibinfo{journal}{\emph{Theoretical issues in ergonomics science}} \bibinfo{volume}{20}, \bibinfo{number}{3} (\bibinfo{year}{2019}), \bibinfo{pages}{284--300}.
\newblock


\bibitem[Becker et~al\mbox{.}(2018)]%
        {becker2018nmc}
\bibfield{author}{\bibinfo{person}{Samantha~Adams Becker}, \bibinfo{person}{Malcolm Brown}, \bibinfo{person}{Eden Dahlstrom}, \bibinfo{person}{Annie Davis}, \bibinfo{person}{Kristi DePaul}, \bibinfo{person}{Veronica Diaz}, {and} \bibinfo{person}{Jeffrey Pomerantz}.} \bibinfo{year}{2018}\natexlab{}.
\newblock \showarticletitle{NMC horizon report: 2018 higher education edition}.
\newblock \bibinfo{journal}{\emph{Louisville, CO: Educause}} (\bibinfo{year}{2018}).
\newblock


\bibitem[Bommasani et~al\mbox{.}(2022)]%
        {60_bommasani2022opportunities}
\bibfield{author}{\bibinfo{person}{Rishi Bommasani}, \bibinfo{person}{Drew~A. Hudson}, \bibinfo{person}{Ehsan Adeli}, {and} \bibinfo{person}{et al.}} \bibinfo{year}{2022}\natexlab{}.
\newblock \bibinfo{title}{On the Opportunities and Risks of Foundation Models}.
\newblock
\newblock
\showeprint[arxiv]{2108.07258}~[cs.LG]


\bibitem[Brown et~al\mbox{.}(2020)]%
        {59_brown2020language}
\bibfield{author}{\bibinfo{person}{Tom~B. Brown}, \bibinfo{person}{Benjamin Mann}, \bibinfo{person}{Nick Ryder}, \bibinfo{person}{Melanie Subbiah}, \bibinfo{person}{Jared Kaplan}, \bibinfo{person}{Prafulla Dhariwal}, \bibinfo{person}{Arvind Neelakantan}, \bibinfo{person}{Pranav Shyam}, \bibinfo{person}{Girish Sastry}, \bibinfo{person}{Amanda Askell}, \bibinfo{person}{Sandhini Agarwal}, \bibinfo{person}{Ariel Herbert-Voss}, \bibinfo{person}{Gretchen Krueger}, \bibinfo{person}{Tom Henighan}, \bibinfo{person}{Rewon Child}, \bibinfo{person}{Aditya Ramesh}, \bibinfo{person}{Daniel~M. Ziegler}, \bibinfo{person}{Jeffrey Wu}, \bibinfo{person}{Clemens Winter}, \bibinfo{person}{Christopher Hesse}, \bibinfo{person}{Mark Chen}, \bibinfo{person}{Eric Sigler}, \bibinfo{person}{Mateusz Litwin}, \bibinfo{person}{Scott Gray}, \bibinfo{person}{Benjamin Chess}, \bibinfo{person}{Jack Clark}, \bibinfo{person}{Christopher Berner}, \bibinfo{person}{Sam McCandlish}, \bibinfo{person}{Alec Radford}, \bibinfo{person}{Ilya Sutskever},
  {and} \bibinfo{person}{Dario Amodei}.} \bibinfo{year}{2020}\natexlab{}.
\newblock \bibinfo{title}{Language Models are Few-Shot Learners}.
\newblock
\newblock
\showeprint[arxiv]{2005.14165}~[cs.CL]


\bibitem[Demir et~al\mbox{.}(2016)]%
        {demir2016team}
\bibfield{author}{\bibinfo{person}{Mustafa Demir}, \bibinfo{person}{Nathan~J McNeese}, {and} \bibinfo{person}{Nancy~J Cooke}.} \bibinfo{year}{2016}\natexlab{}.
\newblock \showarticletitle{Team communication behaviors of the human-automation teaming}. In \bibinfo{booktitle}{\emph{2016 IEEE international multi-disciplinary conference on cognitive methods in situation awareness and decision support (CogSIMA)}}. IEEE, \bibinfo{pages}{28--34}.
\newblock


\bibitem[Driskell et~al\mbox{.}(2006)]%
        {driskell2006makes}
\bibfield{author}{\bibinfo{person}{James~E Driskell}, \bibinfo{person}{Gerald~F Goodwin}, \bibinfo{person}{Eduardo Salas}, {and} \bibinfo{person}{Patrick~Gavan O'Shea}.} \bibinfo{year}{2006}\natexlab{}.
\newblock \showarticletitle{What makes a good team player? Personality and team effectiveness.}
\newblock \bibinfo{journal}{\emph{Group Dynamics: Theory, Research, and Practice}} \bibinfo{volume}{10}, \bibinfo{number}{4} (\bibinfo{year}{2006}), \bibinfo{pages}{249}.
\newblock


\bibitem[Lakhnati et~al\mbox{.}(2023)]%
        {lakhnati2023exploring}
\bibfield{author}{\bibinfo{person}{Younes Lakhnati}, \bibinfo{person}{Max Pascher}, {and} \bibinfo{person}{Jens Gerken}.} \bibinfo{year}{2023}\natexlab{}.
\newblock \showarticletitle{Exploring Large Language Models to Facilitate Variable Autonomy for Human-Robot Teaming}.
\newblock \bibinfo{journal}{\emph{arXiv preprint arXiv:2312.07214}} (\bibinfo{year}{2023}).
\newblock


\bibitem[Manyika and Hsiao(2023)]%
        {manyika2023overview}
\bibfield{author}{\bibinfo{person}{James Manyika} {and} \bibinfo{person}{Sissie Hsiao}.} \bibinfo{year}{2023}\natexlab{}.
\newblock \showarticletitle{An overview of Bard: an early experiment with generative AI}.
\newblock \bibinfo{journal}{\emph{AI. Google Static Documents}}  \bibinfo{volume}{2} (\bibinfo{year}{2023}).
\newblock


\bibitem[McNeese et~al\mbox{.}(2018)]%
        {mcneese2018teaming}
\bibfield{author}{\bibinfo{person}{Nathan~J McNeese}, \bibinfo{person}{Mustafa Demir}, \bibinfo{person}{Nancy~J Cooke}, {and} \bibinfo{person}{Christopher Myers}.} \bibinfo{year}{2018}\natexlab{}.
\newblock \showarticletitle{Teaming with a synthetic teammate: Insights into human-autonomy teaming}.
\newblock \bibinfo{journal}{\emph{Human factors}} \bibinfo{volume}{60}, \bibinfo{number}{2} (\bibinfo{year}{2018}), \bibinfo{pages}{262--273}.
\newblock


\bibitem[OpenAI(2023)]%
        {openai2023gpt4}
\bibfield{author}{\bibinfo{person}{OpenAI}.} \bibinfo{year}{2023}\natexlab{}.
\newblock \bibinfo{title}{GPT-4 Technical Report}.
\newblock
\newblock
\showeprint[arxiv]{2303.08774}~[cs.CL]


\bibitem[Park et~al\mbox{.}(2023)]%
        {park2023generative}
\bibfield{author}{\bibinfo{person}{Joon~Sung Park}, \bibinfo{person}{Joseph O'Brien}, \bibinfo{person}{Carrie~Jun Cai}, \bibinfo{person}{Meredith~Ringel Morris}, \bibinfo{person}{Percy Liang}, {and} \bibinfo{person}{Michael~S Bernstein}.} \bibinfo{year}{2023}\natexlab{}.
\newblock \showarticletitle{Generative agents: Interactive simulacra of human behavior}. In \bibinfo{booktitle}{\emph{Proceedings of the 36th Annual ACM Symposium on User Interface Software and Technology}}. \bibinfo{pages}{1--22}.
\newblock


\bibitem[Prada et~al\mbox{.}(2010)]%
        {prada2010introducing}
\bibfield{author}{\bibinfo{person}{Rui Prada}, \bibinfo{person}{Joao Camilo}, {and} \bibinfo{person}{Maria~Augusta Nunes}.} \bibinfo{year}{2010}\natexlab{}.
\newblock \showarticletitle{Introducing personality into team dynamics}.
\newblock In \bibinfo{booktitle}{\emph{ECAI 2010}}. \bibinfo{publisher}{IOS Press}, \bibinfo{pages}{667--672}.
\newblock


\bibitem[Seeber et~al\mbox{.}(2020)]%
        {seeber2020machines}
\bibfield{author}{\bibinfo{person}{Isabella Seeber}, \bibinfo{person}{Eva Bittner}, \bibinfo{person}{Robert~O Briggs}, \bibinfo{person}{Triparna De~Vreede}, \bibinfo{person}{Gert-Jan De~Vreede}, \bibinfo{person}{Aaron Elkins}, \bibinfo{person}{Ronald Maier}, \bibinfo{person}{Alexander~B Merz}, \bibinfo{person}{Sarah Oeste-Rei{\ss}}, \bibinfo{person}{Nils Randrup}, {et~al\mbox{.}}} \bibinfo{year}{2020}\natexlab{}.
\newblock \showarticletitle{Machines as teammates: A research agenda on AI in team collaboration}.
\newblock \bibinfo{journal}{\emph{Information \& management}} \bibinfo{volume}{57}, \bibinfo{number}{2} (\bibinfo{year}{2020}), \bibinfo{pages}{103174}.
\newblock


\bibitem[Team et~al\mbox{.}(2023)]%
        {team2023gemini}
\bibfield{author}{\bibinfo{person}{Gemini Team}, \bibinfo{person}{Rohan Anil}, \bibinfo{person}{Sebastian Borgeaud}, \bibinfo{person}{Yonghui Wu}, \bibinfo{person}{Jean-Baptiste Alayrac}, \bibinfo{person}{Jiahui Yu}, \bibinfo{person}{Radu Soricut}, \bibinfo{person}{Johan Schalkwyk}, \bibinfo{person}{Andrew~M Dai}, \bibinfo{person}{Anja Hauth}, {et~al\mbox{.}}} \bibinfo{year}{2023}\natexlab{}.
\newblock \showarticletitle{Gemini: a family of highly capable multimodal models}.
\newblock \bibinfo{journal}{\emph{arXiv preprint arXiv:2312.11805}} (\bibinfo{year}{2023}).
\newblock


\bibitem[Zhang et~al\mbox{.}(2023)]%
        {zhang2023investigating}
\bibfield{author}{\bibinfo{person}{Rui Zhang}, \bibinfo{person}{Wen Duan}, \bibinfo{person}{Christopher Flathmann}, \bibinfo{person}{Nathan McNeese}, \bibinfo{person}{Guo Freeman}, {and} \bibinfo{person}{Alyssa Williams}.} \bibinfo{year}{2023}\natexlab{}.
\newblock \showarticletitle{Investigating AI Teammate Communication Strategies and Their Impact in Human-AI Teams for Effective Teamwork}.
\newblock \bibinfo{journal}{\emph{Proceedings of the ACM on Human-Computer Interaction}} \bibinfo{volume}{7}, \bibinfo{number}{CSCW2} (\bibinfo{year}{2023}), \bibinfo{pages}{1--31}.
\newblock


\end{thebibliography}

\end{document}